\documentclass[aps,pra,twocolumn,superscriptaddress,showpacs]{revtex4-2}%
\usepackage{amsfonts}
\usepackage{amsmath}
\usepackage{amssymb}
\usepackage{graphicx}
\usepackage{enumitem}
\usepackage{hyperref}
\usepackage[usenames]{color}
\usepackage[table]{xcolor}
\begin{document}
\title[Vortices in dipolar attractive BECs]{\color{black}  
Vortices in Tunable Dipolar Bose-Einstein condensates with 
Attractive Interactions}
\author{S. Sabari}
\address{
Instituto de Física Teórica, Universidade Estadual Paulista (UNESP), 
01140-070 São Paulo, SP, Brazil}
\author{R. Sasireka}
\address{Department of Physics, Presidency College (Autonomous), Chennai - 600005, India}
\author{R. Radha}
\address{Centre for Nonlinear Science(CeNSc), Department of Physics, Government College for Women(Autonomous), Kumbakonam 612001, India}
\author{A. Uthayakumar}
\address{Department of Physics, Presidency College (Autonomous), Chennai - 600005, India}
\author{Lauro Tomio}
\address{
Instituto de Física Teórica, Universidade Estadual Paulista (UNESP), 
01140-070 São Paulo, SP, Brazil}
\address{Centro Internacional de Física, Instituto de
Física, Universidade de Brasília, 70910-900, Brasília, DF, Brazil
}
\date{\today}

\begin{abstract}
\noindent We investigate the formation of vortices in quasi-two-dimensional  
dipolar Bose-Einstein Condensates (BECs) through the interplay between two-body contact 
and long-ranged dipole-dipole interactions (DDIs), as both 
interactions can be tuned from repulsive to attractive.
By solving the associated Gross-Pitaevskii equation for a rotating system, 
our initial approach concentrates on stabilizing a collapsing condensate with attractive 
$s$-wave two-body interactions by employing sufficiently large repulsive DDIs. 
Subsequently, the same procedure was applied after reversing the signs of both 
interactions to evaluate the sensitivity of vortex formation to such an interchange of interactions.  
As a reference to guide our investigation, valid for generic dipolar atomic species,
we have assumed a condensate with the strong dipolar dysprosium isotope, $^{164}$Dy. 
The correlation of the results with other dipolar BEC systems was exemplified  
by considering rotating BECs with two other isotopes, namely $^{168}$Er and 
$^{52}$Cr. For a purely dipolar condensate (with zero contact interactions) under fixed rotation, we demonstrate how the number of visible vortices increases as the DDI becomes more repulsive, accomplished by tuning the orientation of the dipoles through a characteristic angle parameter.
\end{abstract}

\maketitle

\section{Introduction} Superfluidity, which is believed to be a remarkable macroscopic quantum 
phenomenon, was discovered in the investigation of Liquid Helium by Kapitza and 
Allen~\cite{1938Kapitza,1938Allen}. They showed that liquid He can flow without viscosity below 
a critical temperature. Liquid Helium remained the only bosonic liquid realized in an 
experiment until 1995. The advent of laser cooling with dilute alkali atoms resulted 
in the experimental realization of the new phase of matter called ``Bose-Einstein Condensate" 
(BEC)~\cite{1995Anderson}. The BEC turns out to be a new addition to the family of superfluid, 
which is strikingly different from liquid He, the first member of this family. Liquid Helium is a 
strongly interacting system which makes its theoretical description extremely difficult. 
It has no system parameters like density and interaction strength which can be manipulated 
experimentally. In addition, it has no spin degrees of freedom. On the other hand, almost all 
the parameters like density, kinetic energy and interaction strengths in BECs can be manipulated 
experimentally by engineering the atom laser interaction strength in magnetic or optical traps 
using Feshbach resonance techniques~\cite{2010Chin}. By choosing atomic species and employing 
optical traps to release the spin degrees of freedom, one can realize the spinor 
superfluid~\cite{1998Ho,2002Esslinger} and the dipolar superfluid~\cite{2005Griesmaier}. 
Another advantage of BECs is that ultracold gases are dilute and weakly interacting systems 
which means that they can be easily described by mean-field approximation governed by the usual 
Gross-Pitaevskii (GP) equation or its generalizations.

It should be pointed out that the nonlinearity in a BEC arises by virtue of the two-body short range 
$s-$wave contact interaction. However, realizations of BECs with  long range dipolar atoms
have opened up plenty of avenues leading to new  possibilities in the domain of ultra-cold atoms. Dipolar atoms are  
characterized by having long-range anisotropic interactions with the first BEC realizations
reported with isotopes of Chromium~\cite{2005Griesmaier} followed by  Dysprosium~\cite{Youn:2010} and
Erbium~\cite{Aikawa:2012}. 
Far being different from the isotropic contact interaction in 
condensates with alkali-metal atoms, it is expected that dipole-dipole interaction (DDI) 
can lead to new ground-state properties and interesting phenomena.
Recent experiments have shown that dipolar BECs can form supersolid states 
of matter~\cite{2019Tanzi,2019Bottcher,2019Chomaz}.
Further, dipolar BECs have been endowed with the possibility of playing with two different kinds 
of interactions in a dilute system namely, the conventional two-body contact and the
long-range dipolar interactions, with possibilities to be tuned from attractive to repulsive.
Such dynamical systems associated with BECs have been theoretically and experimentally explored 
in different perspectives, as exemplified by the following 
works and review articles~\cite{2004ODell,Koch:2008,dbec1,2010Lima,2011Muller,dbec2,Sabari2015,
Sabari2018,Sabari2018a,2018Cidrim,2020Diniz,2022Chomaz,Sabari2017,Sabari2019}.
Concerned with dipolar binary mixtures, several investigations have been done, such as
by exploring miscibility properties with different dipolar 
species~\cite{2012Wilson,2017Kumar,2017Kumar-JPC,2018Trautmann,2023Halder}. 
Also, one can trace the actual interest on dipolar mixtures from the recent experimental 
reports on condensation of dipolar molecules~\cite{2024Bigagli}, and on the
production of spin mixtures with Zeeman sublevels of $^{162}$Dy~\cite{2025Lecomte}.

Relevant to the phenomenon of superfluidity, ``quantized vortices" have also 
been intensively investigated before in nondipolar quantum gas experiments, 
as also for binary coupled BECs in \cite{2025kishor,2023daSilva} (and references therein),
with their occurrence in dipolar atoms being reported only recently in
Ref.~\cite{Klaus2022}.
Vortices are considered to be the hallmark of superfluidity representing the breakdown of 
laminar flow with their identification having wider ramifications for understanding 
quantum turbulence~\cite{Feynman}. With the experimental realization of BECs using dipolar atomic isotopes, there has been significant laboratory investigation into vortices in dipolar BECs and their corresponding properties, as recently discussed in~\cite{2023Bland}.
In a BEC, a vortex is described by a solution of the GP equation that carries angular momentum
governed by the geometry and dimensionality of the system under consideration. 
A one-dimensional BEC strictly cannot sustain vortex solution and can support only 
solitons~\cite{2002Khaykovich}. 
While classical vortices can take any 
value of circulation, superfluids are irrotational such that angular rotation is constrained 
to occur through vortices with quantized circulation known as ``quantized vortices".  

In the BECs realized with alkali atoms, the density of the atoms is of the order of $10^4$ atoms/cc. 
At such densities, the interaction between the atoms is weak and predominantly described by $s$-wave 
scattering lengths $a_s$, which being positive indicates repulsive interaction. In such cases, with
$a_s>0$, there is no upper limit to the number of atoms occupying a stable configuration 
which can have more than millions of atoms. This is the situation with 
$^{87}$Rb~\cite{1995Anderson} and $^{23}$Na~\cite{1995Davis}. 
However, if the scattering length becomes negative, the interaction between the atoms is attractive, 
implying that there is  an upper limit to the number of atoms of a stable condensate as verified
experimentally in the case of $^7$Li~\cite{1995Bradley}. 
In addition, it is well understood that vortices are more energetically favored with repulsive 
interactions.
This underscores the fact that the identification of vortices in rotating BECs with 
attractive two-body interactions continues to be 
challenging~\cite{2020Guo,2004Lundh,2004Ghosh,2006Carr}, with their existence
relying on shape deformations of the trap.
Atomic intrinsic properties other than two-body contact interaction can also
provide the necessary stability of a condensate against collapse, as 
shown in Refs. \cite{Sabari2016,2000Gammal,2000GammalJPB,Sabari2022}.

Hence, the main focus of the present work we are reporting 
is to explore the possibility of generating
vortices in dipolar BEC systems having attractive contact interactions with large enough 
repulsive DDI. 
Consistently with the stability analysis performed in 
Ref.~\cite{2002Giovanazzi,2006Griesmaier} within the
Thomas-Fermi limit approximation, in our full numerical approach, we are assuming a 
quasi-2D large trap anisotropy. 
The specific  role of each kind of interaction in vortex production 
is further investigated by inverting the sign of short- and long-range interactions with the stability being accomplished in attractive dipolar systems by the reinforcement of repulsive contact interaction. 

This paper is organized as follows.
In Sect. II, we present the quasi-2D mean-field description to portray the dynamics 
of dipolar BECs with two-body contact interactions. It is pointed out
how one can manipulate experimentally the sign of both interactions independently.
In Sect. III, we present our main results, assuming a hypothetical dipolar BEC,
in two subsections (A and B), by considering the DDI strongly repulsive (A) and 
attractive (B), with the associated contact interactions having opposite signs,
being attractive (or close to zero) in  case (A), and enough repulsive in 
case (B).
In Sect. IV, by assuming three atomic samples for the BECs, we establish a correlation  between the $^{164}$Dy, $^{168}$Er and $^{52}$Cr condensates.
Finally, our main conclusions are highlighted in Sect. V.

\section{The model} 
\begin{figure}[!ht]
\centerline{\includegraphics[width=1.\linewidth]{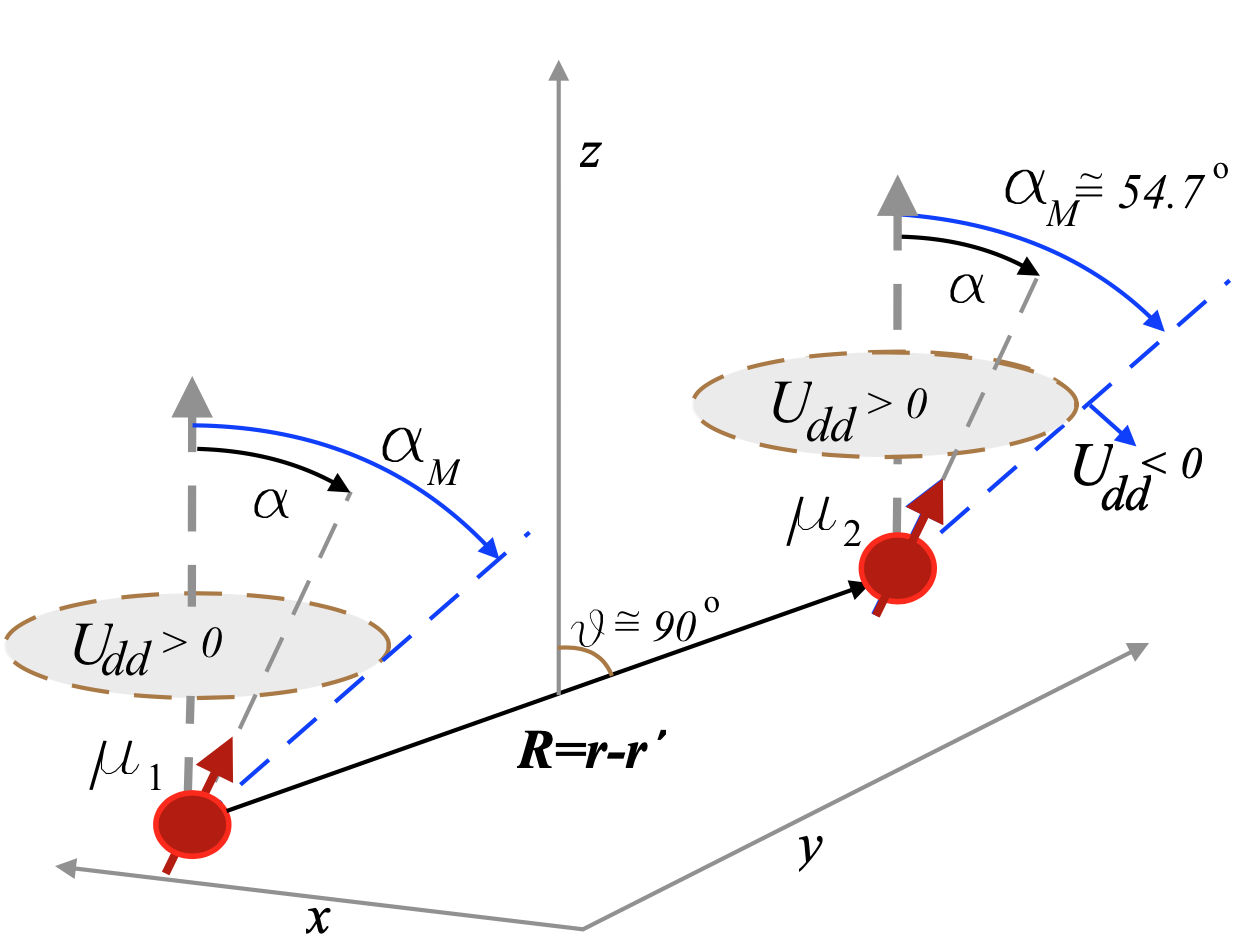}}
\caption{(color online) 
Illustration of the  
dipole–dipole interaction,
as given by \eqref{Udd}, where $\boldsymbol\mu_1=\boldsymbol\mu_2=\boldsymbol\mu$, 
with $\alpha$ being the dipole orientations
related to the $z-$axis which we assume to  tune the 
strength and sign of the effective dipole-dipole interaction.}
\label{rfig01}
\end{figure}

\noindent
A trapped ultracold short-ranged $s-$wave weakly interacting dipolar gas of bosonic 
atoms in a  rotating system can be described by the following three-dimensional (3D) 
time-dependent extension of the GP mean-field formalism~\cite{Koch:2008,dbec1,dbec2}
for the corresponding wave-function $\phi({\mathbf r},t)$: 
{\small\begin{align}
\begin{split}
i\hbar\frac{\partial \phi({\mathbf r},t)}{\partial t}& =\Big[-\frac{\hbar^2}{2m}\nabla^2+V({\mathbf r}) -\Omega L_z + g_s \left\vert \phi({\mathbf r},t)\right\vert^2\\
& +N \int U_{\mathrm{dd}}({\mathbf  r}-{\mathbf r}')\left\vert\phi({\mathbf r}',t)\right\vert^2 d{\mathbf r}' \Big]\phi({\mathbf r},t), \label{eqn:dgpe}
\end{split}
\end{align}
}where $\phi({\mathbf r},t)$ is normalized to one, $\hbar$ is the reduced Planck's 
constant, with $m$ being the atom mass and $N$ the number of total condensed atoms.  
$V({\mathbf r})$ is the trap potential which we assume given by a
non-symmetric 3D harmonic interaction
$V({\mathbf r})=\frac{1}{2} m \omega_\rho^2 [(x^2+y^2)+ 
\lambda z^2],$
where $\omega_x = \omega_y = \omega_\rho$ with 
$\lambda\equiv (\omega_{z}/\omega_{\rho})^2$ being the shape parameter.
With the condensate rotating about the $z-$axis with angular velocity $\Omega$,   
$L_z \equiv -i \hbar (x\frac{\partial}{\partial y} - y \frac{\partial}{\partial x})$ 
will correspond to the z-component of the total angular momentum operator.
The non-linear contact and dipolar interactions are respectively given by the last 
two terms within the square brackets of Eq.~\ref{eqn:dgpe}. 
The strength of the two-body interaction given in terms of the two-body 
scattering length $a_s$, $g_s \equiv 4\pi\hbar^2 a_s N/m$, can be tuned from
attractive (negative) to repulsive (positive) values by employing Feshbach 
resonance techniques~\cite{2010Chin}. 
The long-range interaction between two polarized magnetic dipoles located at 
${\bf r}$ and ${\bf r'}$ as illustrated in Fig.~\ref{rfig01}, 
can be expressed by~\cite{2002Giovanazzi}
{\small\begin{eqnarray}
U_{\mathrm{dd}}({\bf r -r'})=\frac{\mu_0 \mu^2 (1-3\cos^2 \vartheta)}{4\pi}
\frac{(3\cos^2\alpha-1)}{2{{\bf \vert r -r'\vert}^3}},
\label{Udd}
\end{eqnarray} 
}where $\mu_0$ is the permeability of free space, $\vartheta$ is the angle 
between the $z-$axis and the vector position of the dipoles 
${\bf R\equiv r -r'}$,
with $\alpha$ defining the  angle of inclination of both dipole moments
$\boldsymbol{\mu}$ relative to the $z-$direction. 
In our approach, we consider pancake-like confinement with $\lambda\gg 1$ 
such that most of the atoms are close to the transverse plane. Therefore,
the vector position of dipoles ${\bf R}$ can be assumed to be close to 
a plane perpendicular to the $z-$direction or $\vartheta \approx 90^\circ$ 
(with the exact limit being considered in the present 2D case).
For more details, see Refs.~\cite{Sabari2024,2024Tomio} where this formalism 
was considered for vortex production and to study the turbulent dynamics in 
dipolar systems.
Therefore, the angle $\alpha$ describing the orientation 
of the dipoles related to the $z-$axis turns out to be the 
key parameter in Eq.~\ref{Udd} to manipulate and effectively alter 
the DDI. With $\alpha_M\approx 54.7^\circ$ being
the {\it magic angle} at which the DDI vanishes (independently 
of $\mu$), the DDI can vary from repulsive (when $\alpha<\alpha_M$) 
to attractive (when $\alpha>\alpha_M$) interactions.  

For the convenience of using the GP equation (\ref{eqn:dgpe}) in dimensionless form, 
we introduce the dimensionless variables with the length and time units given by
{\color{black}
$\ell_\rho=\sqrt{\hbar/(m \omega_\rho)}$ and $\omega_\rho^{-1}$.
With this procedure, the previous physical quantities are transformed as 
${\bar {\bf r}}= {\bf r}/\ell_\rho,{\bar {\bf R}}={\bf R}/\ell_\rho, \bar t=t\omega_\rho$, 
$ \bar \phi=\ell_\rho^{3/2}\phi$, $\bar\Omega=\omega_\rho^{-1}\Omega$, $\bar V=\hbar\omega_\rho{V}$
with Eq. (\ref{eqn:dgpe}) expressed in terms of the overhead bar variables. By removing the 
overhead bar from all the variables (which will be understood as dimensionless),  
Eq. (\ref{eqn:dgpe}) can be written as 
\begin{align}
\begin{split}
i \frac{\partial \phi({\mathbf{r}},{t})}{\partial t} & = \biggr[-\frac{\nabla^2}{2}+
V({\mathbf r}) -\Omega L_z 
+g_{3D} \vert {\phi({\mathbf{r}},{t})} \vert^2\\
&+ g_{\mathrm{dd}}\int \frac{3\cos^2\alpha-1}{2{\bf \vert r -r'\vert^3}} 
\vert \phi({\mathbf{r}}',t) \vert^2 d{\mathbf{r}}'\biggr] \phi({\mathbf{r}},{t}),\label{gpe3d} 
\end{split}
\end{align}
where $g_{3D}\equiv 4\pi N a_s/\ell_\rho =
g_{s}/ (\hbar\omega_\rho \ell_\rho^3)$
and $g_{\mathrm{dd}}\equiv N\mu_0 \mu^2/(4\pi \hbar\omega_\rho \ell_\rho^3)$.}
In terms of a defined dipole length $a_{dd}\equiv \mu_0\mu^2 m/(12\pi\hbar^2)$,
we can write $g_{dd}=3N(a_{dd}/\ell_\rho)$.
The stability of a dipolar BEC depends on 
the external trap geometry. For example,  a dipolar BEC is stable or unstable 
depending on whether the trap is pancake- or cigar-shaped, respectively. The
instability usually can be overcome by applying a strong pancake trap with 
repulsive two-body contact interaction. The external trap helps to
stabilize the dipolar BEC by imprinting anisotropy onto the density 
distribution. Hence, with the dynamics of the dipolar BEC being 
strongly confined in the axial direction ($\lambda \gg 1$), $\Psi({\bf r},t)$ 
can be decoupled as
{\small \begin{align}\phi({\bf r},t)\equiv 
\chi(z)\psi({\bf {\boldsymbol\rho}},t)\equiv
\left(
\frac{\lambda}{\pi}\right)^{{1}/{4}}
\exp\left(\frac{-\lambda z^2}{2}\right) \psi({\bf {\boldsymbol\rho}},t),
\label{chixyz}
\end{align} 
}where $\boldsymbol{\rho}\equiv(x,y)  \equiv(\rho\cos\vartheta,\rho\sin\vartheta)$
allowing $z$ to be {\color{black} integrated out. 
After simplification, we obtain the effective 2D pancake-shaped dipolar 
BEC~\cite{2006Fisher,2012Wilson} as 
{\small \begin{eqnarray}
 {\rm i}\frac{\partial \psi(\boldsymbol{\rho}, t)}{\partial t}&=&\left\{
 -\frac{1}{2}\nabla_{\rho}^2+\frac{\rho^2}{2}-\Omega L_z +g_{2D}
 |\psi(\boldsymbol{\rho}, t)|^2 \right. \nonumber\\
&+&\left. g_{dd}\int \frac{d^2k_{\rho}}{4\pi^2}
e^{i \mathbf{k}_{\rho}.\boldsymbol{\rho}}\widetilde{n}(\mathbf k_{\rho})
\widetilde{{V}}^{(d)}({\bf k_\rho}) \right\} \psi(\boldsymbol{\rho}, t),
\label{gpe_scaled}
\end{eqnarray}
}where $g_{2D}\equiv\sqrt{8\pi\lambda}(a_s/\ell_\rho)N $.} 
In the above 2D Eq.~\eqref{gpe_scaled}, $\widetilde{n}(\mathbf k_{\rho})$ and $\widetilde{{V}}^{(d)}({\bf k_\rho})$
are the Fourier transforms of the 2D density and dipolar potential respectively. As shown in Refs.~\cite{Sabari2024,2024Tomio}, the DDI in momentum space 
after an averaging of the polarization rotating field in the $(k_x,k_y)$ plane  can be expressed as 
{\small \begin{equation}
{\widetilde{V}}^{(d)}({\bf k_\rho}) = \frac{3\cos^2\alpha-1}{2}
\left[2-3\sqrt{\frac{\pi}{2\lambda}}k_\rho
e^{\frac{k_{\rho }^{2}}{2\lambda}} 
{\rm erfc}\left( \frac{k_{\rho }}{\sqrt{2\lambda}}\right)\right] 
\label{DDI-2D}
\end{equation} 
}where ${\rm erfc}(x)$ is the complementary error function of $x$.
Therefore, as shown above, the strength of the atom-atom long-ranged dipolar interaction 
is guided by the factor $g_{dd}\left(3\cos^2\alpha-1\right)/2$
which varies from $g_{dd}$ for $\alpha=0^\circ$ to $-{g_{dd}}/{2}$ for $\alpha=90^\circ$
while it vanishes for $\alpha\approx54.7^\circ$. 
In other words, by changing the orientation of the dipoles with   $z$-axis, 
one can change not only the magnitude of DDI but also the direction changing the  interaction from
repulsive to attractive as well. So, restricted by possible stability 
requirements, the two kinds of nonlinear interactions 
present in  equation \eqref{gpe_scaled} (short- and long-ranged ones), in principle, can be 
manipulated independently: 
In other words, for the contact interaction, we have the Feshbach techniques to change $a_s$ from attractive to
repulsive  and for the DDI (as shown in Fig.\ref{rfig01}),
we have the orientations of the two dipoles defined by the angle $\alpha$ 
(once considered $\vartheta$ fixed at $90^\circ$).
In terms of $\alpha$, we can assume the DDI strength to be redefined as 
\begin{equation}
a_{dd}(\alpha)= a_{dd} \frac{3\cos^2\alpha-1}{2},\label{add-alpha}
\end{equation}
so that it can vary from
maximum repulsive [$a_{dd}(0^\circ)=a_{dd}$] to maximum attractive [$a_{dd}(90^\circ)=-a_{dd}/2$].

For the numerical solution of  Eq.~(\ref{gpe_scaled}) to investigate the vortex formation 
in the condensate, we have combined the usual split-step Crank-Nicholson method together 
with the fast Fourier transform approach. In our numerics, we have used a $512 \times 512$ 
grid size with $\Delta x=\Delta y=0.04$ for both $x$ and $y$ (units $\ell_\rho$) with time 
step $\Delta t=0.01$ (units $\omega_\rho^{-1}$). Assuming initially the strong dipolar 
Dysprosium isotope $^{164}$Dy as the hypothetical atomic sample in our investigation, 
we kept the number of atoms fixed at  $N=1.0 \times 10^3$ with the  rotational 
frequency being $\Omega=0.9$ 
(units $\omega_\rho$) and the length unit such that
$\ell_\rho=1\times10^{-6}m$.
In this case, $a_{dd}\approx 6.94\times 10^{-3} \ell_{\rho}\approx 131\,a_0$, where 
$a_0=0.52918\times 10^{-10} m$ is the Bohr radius. In order to compare with other
dipolar BECs made up of Erbium and Chromium isotopes, the long-range DDI can be kept at 
$a_{dd} \approx 66\,a_0$ ($^{168}$Er) and $a_{dd} \approx 16\,a_0$ ($^{52}$Cr) respectively. For all vortex studies, we first prepare a stable initial state using imaginary time propagation. Subsequently, by introducing the rotational frequency $\Omega$, we evolve the system in real time by using the split-step Crank-Nicholson method.

\section{Vortex formation in dipolar BECs with Attractive interactions}
In this section, we study the interplay between long- and short-range interactions with
one of these interactions being attractive. For that, we have numerically analyzed the
corresponding energetic parameters and rotational properties of the condensates. In part (A), our primary objective was to stabilize a collapsing condensate with attractive $s$-wave interactions using a sufficiently large repulsive DDI. This was followed by an analysis of vortex generation at a rotational frequency $\Omega = 0.9$, which is close to the transverse trap frequency.
In part (B), we reversed the roles of the long- and short-range interactions, considering an attractive DDI. Starting with a non-rotating, stable BEC system, we subjected it to high rotation ($\Omega = 0.9$) and analyzed the sensitivity of vortex formation on the nature of the interactions.

\subsection{ Attractive $s$-wave two-body interaction}
\begin{figure}[!ht]\centerline{\includegraphics*[width=1.0\linewidth]{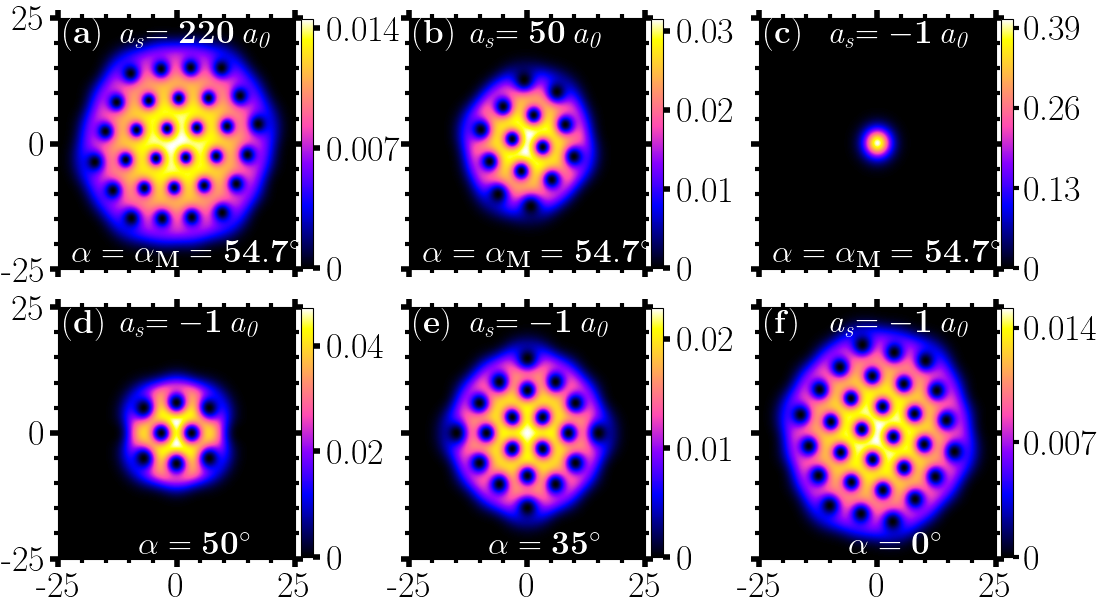}}
\caption{(Color online) Vortex patterns in ($x,y$) density distributions given by
$\Omega=0.9$ (unit $\omega_\rho$) for a dipolar BEC in which $\alpha$ is the dipolar parameter 
controlling the strength as given by  equation \eqref{add-alpha}. In the upper panels, 
the DDI is set to zero by $\alpha=\alpha_M=54.7^\circ$ with the contact 
interactions varying from repulsive, $a_s=220a_0$ (a) and $a_s=50a_0$ (b), 
to attractive $a_s=-1a_0$ (c). 
In the bottom panels, the contact interactions are kept attractive as in (c) 
with DDI being increasingly repulsive with $\alpha$ varying as 
$\alpha=50^\circ$ (d), $\alpha=35^\circ$ (e), and $\alpha=0^\circ$ (f).
The density levels are shown in bars at the right of each panel with
the length unit being $\ell_\rho$.
}
\label{rfig02}
\end{figure}
The possibility of generating visible vortices in attractive BECs 
has been studied before for non-dipolar systems with  
deformed trap configurations. Even without rotation, the condensate stability 
is limited up to some critical number of atoms. By considering dipolar systems,
our aim is to explore the vortex production while having attractive contact 
interaction which will be compensated with enough repulsive dipolar interactions.
To highlight the impact of attractive short-range $s$-wave interaction
on the vortex production, we first switch off completely the long-range DDI 
and consider the reduction of the $s$-wave two-body repulsive interaction. The results
shown for the densities $\left\vert \psi \right\vert^{2}$ are presented in the upper 
panels of Fig.\ref{rfig02}.
Represented in panels (a), (b) and (c), we have $a_s$ being reduced from 
$a_s=220a_0$ (panel (a)) till $a_s=-1 a_0$ (panel (c)).
As shown in the previous section, this is realized in our approach by fixing 
the dipolar angle $\alpha$ at $\sim 54.7^\circ$ such that the dipolar condensate
will behave like  a non-dipolar system. As a reference atomic sample, we  consider
a $^{164}$Dy BEC for our investigation.
In panel (c), both interactions are close to zero with the
nonlinearity still being  maintained by a small negative $a_s$ which means that the 
binary interaction becomes attractive. In such a case, the condensate can exist 
only for a critical number of atoms~\cite{Sabari2015,Sabari2016,Sabari2018} before it 
collapses. 
Here, we use a fixed rotational frequency with $\Omega=0.9$, large enough to generate visible 
vortices in condensates having enough distribution of the density within the trap. 
Actually, in a BEC with a rotating trap potential, visible vortices begin to appear
only at a certain rotational frequency called critical frequency
($\Omega_c$)~\cite{Sabari2017,Sabari2019}. This implies that in BECs confined 
to a rotating trapping potential $\Omega<\Omega_c$, ghost vortices are first generated 
in the low-density region of the condensates and they do not carry any 
significant angular momentum. 
Also, at $\Omega = \Omega_c $, the ghost vortices start to enter the density 
region of the BECs and get transformed into visible vortices. 
From the results in panel (a) of Fig.~\ref{rfig02}, we see how visible vortices 
appear in the absence of DDI for repulsive binary interaction at $a_s=220a_0$.   
By diminishing the strength of the short-ranged repulsive interaction, the number 
of visible vortices continuously decreases [panel (b) of Fig.\ref{rfig02}],  
while the density increases. This process continues till  
the area becomes too small to contain (or generate) a single vortex.
At this juncture, by changing the binary contact 
interaction from repulsive to attractive, we end up with a situation where the density 
becomes so high and the area shrinks that no vortices are allowed even with higher rotational 
frequencies. This is the general expected behavior for a non-dipolar system in which the
nonlinear contact interactions play the main role. However, considering a dipolar system
in which the dipolar strength can be controlled through the angle  $\alpha$,
as given by \eqref{add-alpha}, an obvious strategy is  to generate vortices in the condensate
by increasing the repulsive long-range interaction using $\alpha$ as a tuning parameter.
\begin{figure}[!ht]
\centering \includegraphics[width=1.0\linewidth]{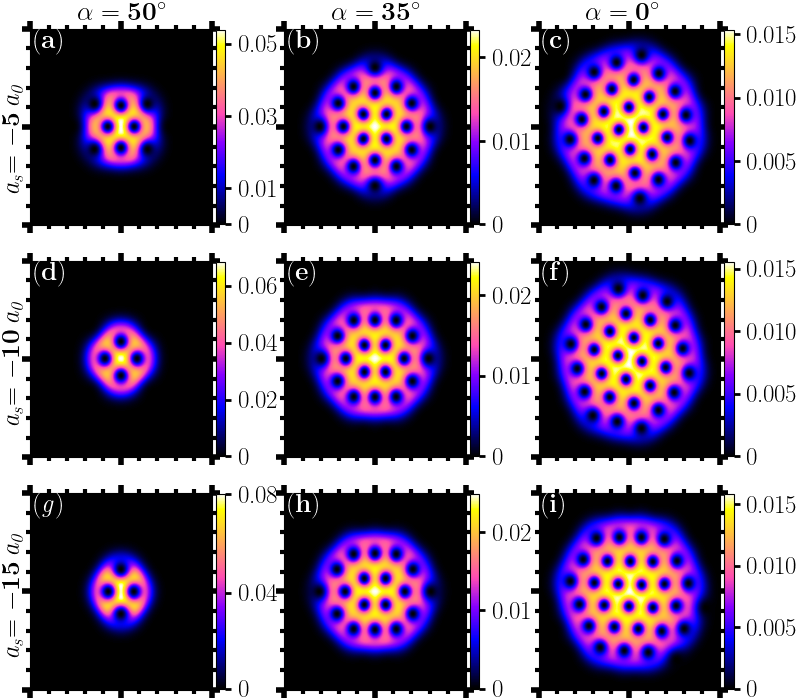}
\caption{\color{black}(Color online) With the same rotation, panel sizes,
and units as in Fig.~\ref{rfig02}, here the vortex patterns are
for a dipolar BEC with attractive contact interactions indicated 
at the left of the row sets ($a_s=-5a_0, -10a_0, -15a_0$).
The DDI strengths (identified by $\alpha$) are fixed for each column with 
the maximum repulsive being $a_{dd}(0^\circ)$ ($=131a_0$ for $^{164}$Dy).
}
\label{rfig03}
\end{figure}
This is exemplified in panels (d), (e) and (f) of Fig.~\ref{rfig02} with
$\alpha$ given by $50^\circ$, $35^\circ$, and $0^\circ$, respectively.
The sensitivity of the dipolar interactions as related to the contact interactions can
already be verified by comparing the upper panels (when DDI is set to zero) with the lower ones
(when the contact interaction is attractive close to zero). Notice that just by changing
about 4.7 degrees of the parameter $\alpha$, the dipolar condensate goes from zero vortices to
display about 8 vortices, as verified by going from panels (c) to (d).
In this regard, another point to be noticed  is that the increase in the number of vortices for 
pure repulsive contact interactions which  goes almost linearly from 0 to about 30 with $a_s$ 
varying from 0 $a_0$ (c) to 220 $a_0$ (a)
whereas in the case of pure dipolar case, this happens for $\alpha$ decreasing from 
54.7$^\circ$ (c) to 0$^\circ$ (f).
\begin{figure}[!ht]
\centerline{\includegraphics*[width=9cm]{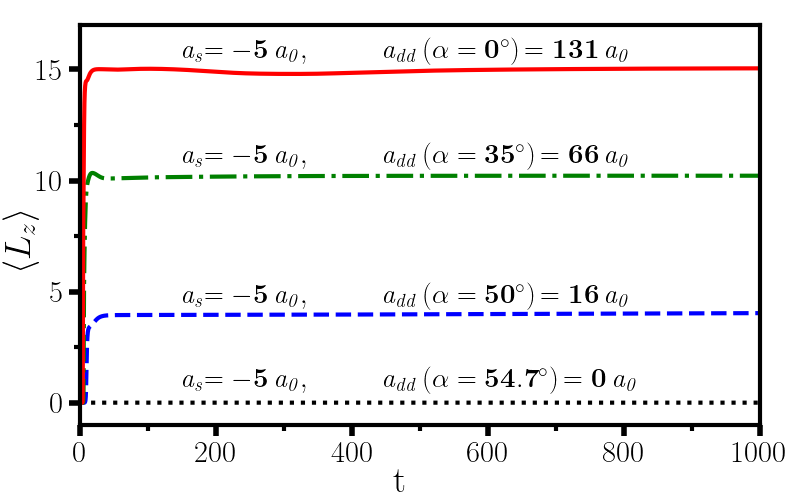}}
\caption{\color{black}(Color online) Time evolution in units 
$\omega_\rho^{-1}$ of the averaged angular momentum per particle
(units $\hbar$),
along the vortex formation for $\Omega=0.9$ (units $\omega_\rho$) and
attractive contact interaction $a_s=-5a_0$ considering
four different cases. To have maximum DDI, we  assume the BEC to be made up of  
$^{164}$Dy.
}
\label{rfig04}
\end{figure}

Hence, to demonstrate the vortex formation in the condensate with attractive two-body interaction, 
we start with the case in which the contact interaction is slightly negative
with $a_s=-1a_0$, but with the DDI being repulsive, which is shown in panel (d) of 
Fig.\ref{rfig02} with $\alpha=50^\circ$.
Next, as expected, more visible vortices are observed as we increase the repulsive DDI by 
decreasing the value of $\alpha$  with 
panels (e) and (f) showing the density distributions for $\alpha=35^\circ$ and $\alpha=0^\circ$, 
respectively.
Further, it is worth mentioning that the results for $^{164}$Dy BEC with $\alpha=50^\circ$ 
and $\alpha=35^\circ$ are almost equal to the $^{52}$Cr and $^{168}$Er BECs respectively 
in their maximum DDI configurations once the masses and transversal trap frequencies are 
appropriately rescaled to fix the length unit
(more details are provided at the end of this section and in Sect. IV).  

Next, to verify the relative sensitivity in the results of the vortex pattern  with changes provided 
by attractive contact interactions related to variations of the repulsive DDIs 
in Fig.\ref{rfig03}, we have a set of panels with density results
obtained with fixed $\Omega=0.9$. At each row, the attractive contact interaction is
fixed such that $a_s=-5a_0$ (top row), $a_s=-10a_0$ (middle row), and
$a_s=-15a_0$ (bottom row) with the DDIs varying for each column of panels with 
$\alpha=50^\circ$ (left), $\alpha=35^\circ$ (middle). and $\alpha=0^\circ$ (right).
It can be observed that  once we fix the DDI,  change of  the contact interactions from $-5a_0$ 
to $-15a_0$ does not have  any impact   on the number of vortices  and is more  pronounced only  for the
case  when the repulsive DDI is not large (as in case that $\alpha=50^\circ$). 
When the DDI is large enough, as in the second and third columns, we have almost 
no changes in the number of visible vortices for a  variation of about 10$a_0$ 
in the repulsive contact interactions.

Further, in Fig.~\ref{rfig04}, the stability of the generated vortices is examined 
by fixing the attractive short-range interaction at $a_s=-5a_0$ and varying the strengths of the DDI. 
This is achieved by considering different dipole orientation angles: $\alpha=0^\circ$ (maximum DDI), 
$\alpha=35^\circ$, $\alpha=50^\circ$, and $\alpha=54.7^\circ$ (zero DDI).
The average angular momentum which gives a measure of stability  is found to be  maximum for a $^{164}$Dy BEC such that 
$a_{dd}= 131a_0$. The other two values for $\alpha=35^\circ$ and $\alpha=50^\circ$
refer to the maximum DDI of $^{168}$Er and $^{52}$Cr with $a_{dd}=66a_0$ 
and $a_{dd}=16a_0$ respectively.
The stability of the produced vortices is probed by considering the time evolution of
the averaged angular momentum per particle, $\langle L_z\rangle$ with the saturation of 
$\langle L_z\rangle$ representing the stability. 
From Fig.\ref{rfig04}, one observes that $\langle L_z\rangle$ for 
$\alpha=0^\circ$ is found to be maximum compared to that of $\alpha=35^\circ$ 
and $\alpha=50^\circ$. When ${\alpha=54.7^\circ}$, we have pure attractive contact 
interaction with no DDI, implying zero $\langle L_z\rangle$.

\subsection{Attractive dipole-dipole interaction}
\begin{figure}[!ht]
\centering \includegraphics[width=1\linewidth]{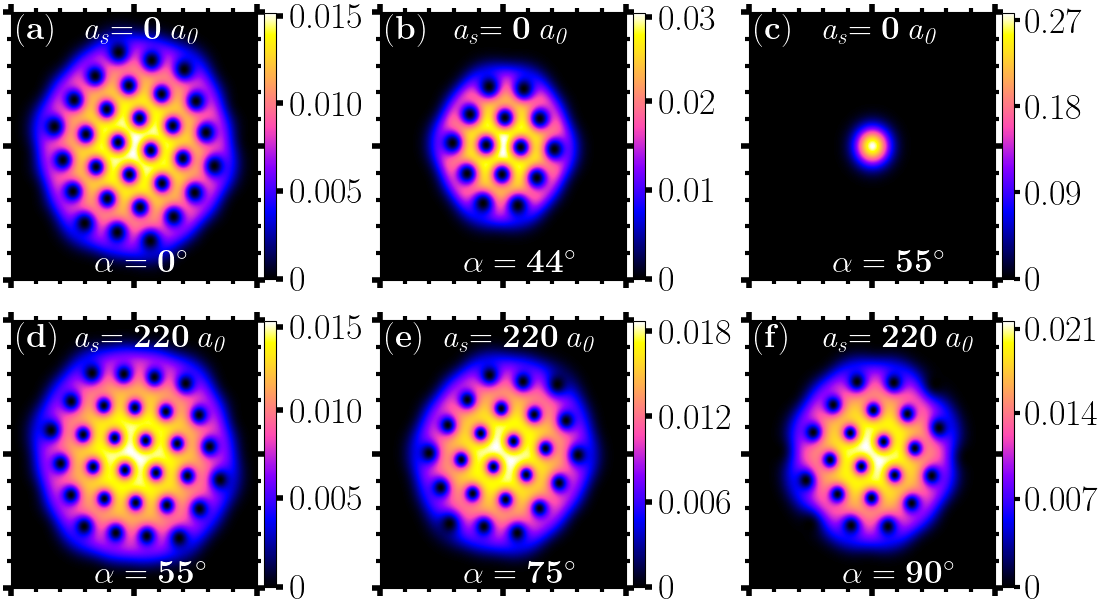}
\caption{(Color online) 
With the same rotation, panel sizes, and units as in 
Figs.~\ref{rfig02} and
\ref{rfig03}, here the vortex patterns are shown for
dipolar $^{164}$Dy BEC considering 
($i$) $a_s=0$ (pure dipolar system) in the upper row
with a repulsive DDI going from its maximum ($\alpha=0$) to zero ($\alpha=55^{\circ}$); and 
($ii$) $a_s=220 a_0$, in the lower row, considering 
only attractive DDI, from slightly negative(attractive) ($\alpha=55^\circ$) 
to maximum attractive ($\alpha=90^{\circ}$). 
}
\label{rfig05}
\end{figure}

Our main focus in this subsection is to study the formation of vortices in 
the condensate in the presence of an attractive DDI. For that, we present
in Fig.~\ref{rfig05} the  results for density profiles  
$\left\vert \psi \right\vert^{2}$ of the dipolar $^{164}$Dy BEC.
In the upper row of Fig.~\ref{rfig05}, 
we first turn off the $s$-wave two-body interaction ($a_s=0$)
to highlight the impact of the DDI when being changed from 
repulsive to attractive  fixing  the angular velocity of rotation such that $\Omega=0.9$.
The repulsive DDI is kept maximimum  in panel (a) with $\alpha=0$ such that $a_{dd}(0^\circ)=a_{dd}=
131a_0$ for the $^{164}$Dy showing large number of vortices. When we increase the angle $\alpha=44^\circ$(reduce the magnitude of repulsive DDI) ,the number of stable vortices decreases (shown in panel(b)).  While the DDI  is  close to zero in panel (c) with $\alpha=55^\circ$,
  the density collapses to such an extent that   a single vortex can no longer exist.
Next, in the lower row of this figure, 
we start in panel (d) by fixing the DDI close to zero as in (c)  and adjust the
repulsive contact interaction to reproduce the vortex pattern formation
previously obtained in panel (a). This is done to primarily to  establish a 
correlation  between long range  and short-range interactions in
the visual production of vortices in the condensates. Starting with panel (d), we have the panels 
(e) and (f) to show  how far we can reduce the number of vortices by 
increasing the attractive long-range interaction by changing the angle $\alpha=90^\circ$  for the $^{164}$Dy condensate. It can be observed from panels (d-f) that despite introducing maximum attractive DDI, we are able to sustain vortices.

\begin{figure}[!ht]
\centering \includegraphics[width=1\linewidth]{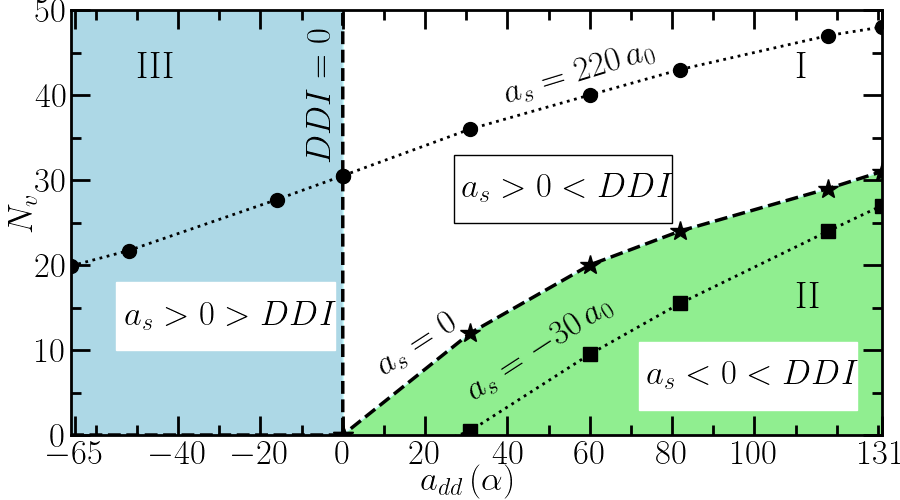}
\caption{(Color online) Number of visible vortices ($N_v$) with
rotation being fixed at $\Omega=0.9$ (units $\omega_\rho$) is 
plotted against DDI given by \eqref{add-alpha} (in units of $
a_0$). The results are shown for three different values of the
contact interaction strengths: $a_s=$-30$a_0$ (squares connected
by a dotted line), 0 (starts connected by a dashed line), and 
220$a_0$ (bullets connected by a dotted line). The 
symbols represent visually identified results (within a
possible error of $\pm 1$). Explicitly indicated inside the
figure, the colors identify three regions where vortex production
can be verified, which are separated by dashed lines.
}\label{rfig06}
\end{figure}%
\subsection{Correlation between long-range DDI and
contact interactions on the vortex production}

In this section, in Fig.~\ref{rfig06}, we present a summary of our results 
highlighting the number of visible vortices ($N_v$) that emerge under 
a fixed rotation ($\Omega = 0.9$) as we vary the strengths of both 
dipole-dipole interactions (DDI) and contact interactions from attractive to repulsive domain.
This occurs as long as the interactions are suitably balanced ensuring that the net atom-atom interaction remains repulsive. In the results shown in Fig.~\ref{rfig06}, we consider a condensate of $^{164}$Dy as a representative sample as it is the isotope with the largest dipole-dipole interaction (DDI) used in Bose-Einstein condensate (BEC) experiments with a maximum DDI given by $a_{dd} = 131 a_0$. Notably, we observe a correlation between contact and long-range interactions in vortex production when one of them is tuned to zero and the other is sufficiently repulsive. For example, with $a_s = 0$, the number of visible vortices is approximately $N_v \approx 30$ for $a_{dd}(0^\circ) = 131 a_0$ (maximum DDI). Similarly, when $a_{dd} = 0$, $N_v \approx 30$ for $a_s = 220 a_0$.

The vortex production is directly correlated with the corresponding root-mean-square (rms) radius. 
Results for the number of vortices ($N_v$) as a function of the DDI are explicitly presented in Fig.~\ref{rfig06}, 
considering three fixed strengths  of the contact interaction . 
For attractive contact interactions exemplified by $a_s = -30 a_0$, 
the results are represented by box symbols connected with a dotted line. 
In the case of zero contact interactions ($a_s = 0$), the results are shown as blue star symbols connected by a dashed line. 
Lastly, for repulsive contact interactions ($a_s = 220 a_0$), the results are depicted as bullets connected by a solid line.
This large repulsive contact interaction was chosen to establish a 
direct correspondence in the number of visible vortices, $N_v \approx 30$, 
between the scenario where $a_{dd}(\alpha) = a_{dd}(54.7^\circ) = 0$ and 
the case where the DDI is at its maximum value, $a_{dd}(\alpha) = a_{dd}(0^\circ) = 131 a_0$.

However, this representation can be considered universal 
as appropriately rescaling the $x$-axis along with the corresponding $N_v$ allows 
one to describe vortex production for other dipolar species such as $^{168}$Er or $^{52}$Cr. 
These species have maximum DDIs of $a_{dd}(0^\circ) = 66 a_0$ (with $N_v \approx 20$) 
and $a_{dd}(0^\circ) = 16 a_0$ (with $N_v \approx 8$), respectively. 
Therefore, to generalize the results shown in Fig.~\ref{rfig06}, the diagrams must be rescaled accordingly.

We also find it convenient to identify three distinct regions in the vortex production results, 
separated by two main dashed lines in the figure (for $a_{dd} = 0$ and $a_s = 0$). 
Region I (white) corresponds to the case where both interactions are repulsive. 
Region II (greenish) represents the scenario where the contact interactions 
are attractive while the DDI remains repulsive. Lastly, Region III (bluish) 
denotes the case where the DDI is attractive, paired with positive (repulsive) contact interactions.
To evaluate the relative impact of short-range $s$-wave and long-range DDIs on vortex production, 
we analyze the variation in the number of vortices, $\Delta N_v$, 
over a given interval of $a_{dd}(\alpha)$. For example, in Fig.~\ref{rfig06}, 
we consider the interval $a_{dd}(\alpha) = 30a_0$ to $131a_0$, 
which corresponds to the DDI range in region II where $a_s = -30a_0$. 
In this case, we observe $\Delta N_v = 27$. Using the same interval, 
$\Delta N_v = 15$ for $a_s = 0$  and $\Delta N_v = 13$ for $a_s = 220a_0$. 
These results clearly demonstrate that repulsive DDIs have a more pronounced 
effect on vortex production when the contact interactions are attractive.
These results indicate that turbulence could be more easily reached 
by dynamical variations of repulsive DDIs with negative contact interactions.

\section{Correlation between different dipolar condensates}
In this section, we attempt to establish a correlation between BECs arising out of the three dipolar atoms that have been considered in cold atom experiments ($^{164}$Dy, $^{168}$Er, and $^{52}$Cr)
with reference to the physically observable quantities like energy, chemical potential, vortex number, and rms radius.   
Using the $^{164}$Dy dipolar BEC as the reference system, we adjust the corresponding orientation of its 
dipoles relative to the $z$-axis as given by the angle $\alpha$. This adjustment allows the DDI of $^{164}$Dy 
to be tuned to match with that of the $^{168}$Er and $^{52}$Cr BECs with respect to measurable quantities 
such as energy, chemical potential, and rms radius.
Here, we primarily focus on matching the number of visible vortices to establish a correlation between two different condensates. 
In this way, by considering $^{164}$Dy
as the reference dipolar atomic system and adjusting its  DDI through the parameter $\alpha$ appropriately, the same
number of vortices of other dipolar atoms like Erbium and Chromium can be obtained. 
For example, let us consider $^{52}$Cr being the other atomic species. In this case,
we need to tune $\alpha$ such that 
$a_{dd}(\alpha_{Dy})=a_{dd}(\alpha)=a_{dd}(\alpha_{Cr}=0)=16a_0$. In other words, Dysprosium at $\alpha=50^\circ$
is correlated to Chromium at $\alpha=0^\circ$.
Using this procedure, the correlation between the observables of the three dipolar atoms in the vortex production
is been summarized in Table~\ref{tab:1} for pure dipolar condensates (setting $a_s$ to be zero) of  $^{164}$Dy, $^{168}$Er, and $^{52}$Cr.

\begin{figure}[!ht]
\centering \includegraphics[width=8.5cm,height=6cm]{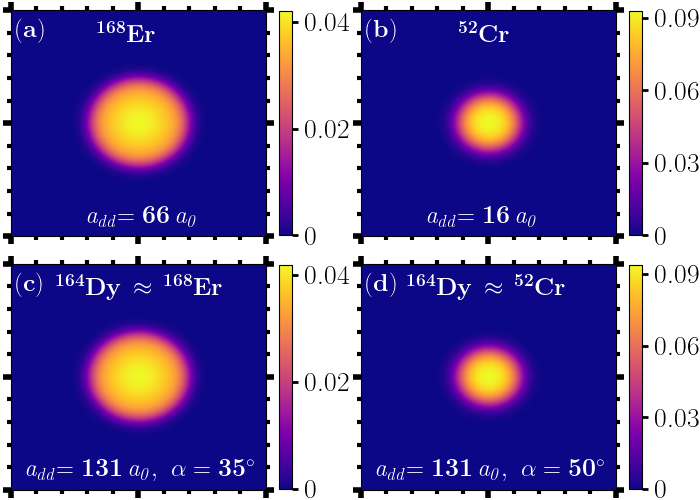} 
\caption{\color{black}(Color online) With no rotation ($\Omega=0$), zero contact interactions ($a_s=0$) and maximum DDIs ($\alpha=0$, in the corresponding independent unit system), in the upper panels, we show the densities 
of the dipolar BEC systems $^{168}$Er (a) and $^{52}$Cr (b). 
They are respectively compared with $^{164}$Dy BEC densities, in (d) and (e), by matching their rms radius 
obtained after tuning the respective dipole orientations.
For that, $\alpha=35^\circ$ in (c) to obtain a correlation with the density of $^{168}$Er (a); whereas 
$\alpha=50^\circ$ in (d) to correlate  with the density of
$^{52}$Cr. The density levels are indicated in bars at the right of the panels with length scales according to the
respective atomic condensate that is being considered.}
\label{rfig07}
\end{figure}

\begin{figure}[!ht]
\centering \includegraphics[width=8.5cm,height=6.5cm]{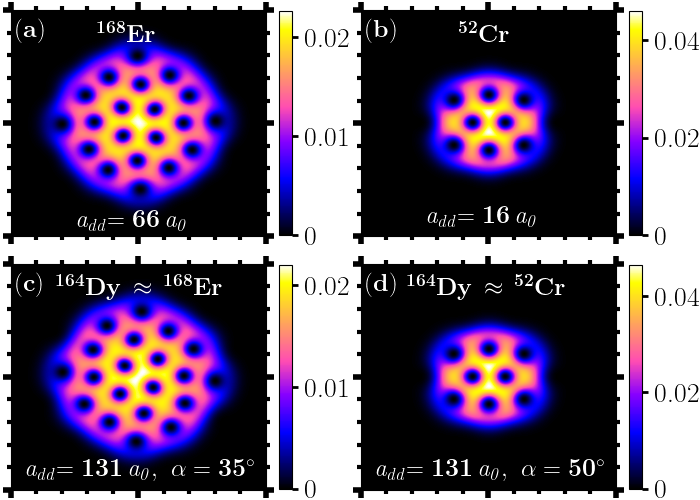}
\caption{(Color online) Following the same as in Fig.~\ref{rfig07}, but
introducing a rotational frequency $\Omega=0.9$, in  units of the transversal
trap frequency considered for each specific condensate.
}
\label{rfig08}
\end{figure}
\begin{table}[!ht] 
\center
\caption{Physical quantities and interaction parameters as considered in 
Figs.~\ref{rfig07} (no rotation) and~\ref{rfig08}
(with rotation given by
$\Omega=0.9$) together with the corresponding numerical results. In the first three 
rows, as indicated, the results are for 
$^{164}$Dy, $^{168}$Er and $^{52}$Cr when maximizing their DDIs ($\alpha=0$, within their specific 
independent units). In the  4th and 5th rows, the results are for $^{164}$Dy with $\alpha$ 
tuned to reproduce approximately the same values of the rms radius previously obtained for 
$^{168}$Er and $^{52}$Cr respectively(corresponding to the similar vortex patterns presented in 
Fig.~\ref{rfig08}).
}
\label{tab:1} 
\begin{tabular}{c|cccccccc}
\hline\hline BECs&$\alpha$&$\displaystyle\frac{a_{dd}(\alpha)}{a_0}$&$\mu (\hbar\omega_\rho)$ & $E (\hbar\omega_\rho)$ &$\sqrt{\langle \rho^2\rangle}(\ell_\rho)$ & \;\;\;$N_v$\\  \hline \hline
$^{164}$Dy &$0^\circ$  &131&6.47 & 4.14 & 2.58 & 30 \\  
$^{168}$Er &$0^\circ$  &66&4.62 & 2.98 & 2.16 & 20  \\  
$^{52}$Cr &$0^\circ$ &16&2.35 & 1.60 & 1.53 &  8 \\  \hline 
$^{164}$Dy &$35^\circ$ &66& 4.63 & 2.98 & 2.16 &20 \\  
$^{164}$Dy &$50^\circ$ & 16&2.33 & 1.58 & 1.52 &8  \\  
\hline\hline
\end{tabular}
\end{table}

The matching of the number of visible vortices $N_v$ is visually identified by considering
the density plots (in which we are ignoring vortices hidden in the low-density regions). 
This procedure can address any small deviation found in the other observables like energy 
and chemical potential. As shown, the number of visible vortices 
is directly proportional to the rms radius. By tuning the dipolar angle of the 
$^{164}$Dy BEC, such that $\alpha \approx 50^\circ$, its rms radius will be reduced 
to match with that of a BEC with $^{52}$Cr keeping its DDI to be maximum 
(ie., $\alpha=0^\circ$).
Similarly, by adjusting the orientation of the dipoles such that $\alpha = 35^\circ$, 
the DDI of $^{164}$Dy can be tuned to that of $^{168}$Er condensate keeping its DDI to be maximum (ie.$\alpha=0^\circ$) so  that the number of visible vortices and the rms radius 
exactly match for both atomic systems. 

Related to the results shown in Table~\ref{tab:1} for $a_s=0$, we also have the 
density plots shown in Figs.~\ref{rfig07} and \ref{rfig08}, 
in which we are comparing the corresponding sizes of the dipolar condensates 
as well as the vorticity.
In Fig.~\ref{rfig07}, without rotation, the condensate densities obtained
for $^{168}$Er (a) and $^{52}$Cr (b) keeping their  DDI to be maximum, 
are being compared with the densities obtained from $^{164}$Dy
in (c) with $\alpha=35^\circ$ and (d) with $\alpha=50^\circ$ respectively. 
In Fig.\ref{rfig08}, we present the same four density panels with the 
rotational frequency given by $\Omega=0.9$. 
Therefore, from these results, for pure dipolar systems, it is evident that by 
appropriately adjusting the orientation of the dipoles relative to the $z$-axis, 
one observes perfect correlation between the  density profiles of  $^{164}$Dy BEC  with that  of  $^{168}$Er and 
$^{52}$Cr BECs.

\section{Conclusion}
In this paper, we explore the generation of energetically stable 
vortices in a quasi-two-dimensional dipolar BEC endowed with both short- and 
long-range interactions (contact $s$-wave and DDIs, respectively). 
We consider a stable configuration with a fixed high rotation close to 
the transversal trap frequency, $\Omega = 0.9$ (in units of $\omega_\rho$). 
To achieve this, we first assume an attractive contact $s$-wave interaction ($a_s < 0$) 
and tune the long-range DDI to be sufficiently repulsive by 
varying the dipole orientations maximizing the DDI when both dipoles are 
aligned along the $z$-axis (i.e., $\alpha = 0$). Furthermore, by reversing 
the roles of the attractive and repulsive interactions, we establish a stable 
rotating configuration with attractive DDI (for $\alpha > 54.7^\circ$) and 
repulsive contact interactions ($a_s > 0$). Using the $^{164}$Dy isotope 
as a reference sample, we investigate the interplay between the long-range DDI 
and short-range $s$-wave interactions 
analyzing their relative equivalence in the formation of vortices.

A key result of our study on the vorticity of a quasi-2D dipolar BEC is 
the determination of the maximum number of visible vortices achievable 
for a given contact interaction under high rotation near the transverse 
rotational frequency of the trap. Our findings show that for a rotation of 
$\Omega = 0.9$ and zero contact interaction, the maximum number of 
vortices for $^{164}$Dy (with $a_{dd} = 131 a_0$) is approximately $N_v \approx 30$. 
This number can only increase if the DDI exceeds this value, as shown in 
Fig.~\ref{rfig06}. Additionally, we observe that vortex production is more 
pronounced for attractive contact interactions ($a_s < 0$) than for repulsive 
ones ($a_s > 0$) providing insight into the control and study of quantum turbulence in dipolar BECs.

We have also explored the relationship between dipolar atomic BECs of different 
species for vortex formation comparing $^{164}$Dy with $^{168}$Er and $^{52}$Cr. 
By manipulating the orientation of the dipoles, we studied the effects on 
observable quantities such as the rms radius, energy, and chemical potential, 
(all calculated for zero rotation). Under the same conditions as for $^{164}$Dy, 
with $a_s = 0$ and $\Omega = 0.9$ (in units of the respective transverse trap frequencies), 
the maximum number of vortices was found to be $N_v \approx 20$ for $^{168}$Er and $N_v \approx 8$ for $^{52}$Cr.

Our results underscore that vortices can be generated in systems with 
short-range attractive interactions, provided there is sufficient repulsive long-range DDI. 
Conversely, vortices can also be produced when the roles of the long- and 
short-range interactions are reversed with attractive DDIs balanced by 
sufficiently large repulsive contact interactions. Since the findings of this 
paper are applicable to other dipolar atomic species, we believe this work 
offers a valuable foundation for both theoretical and experimental studies 
involving rotating dipolar BECs and the exploration of suitable combinations 
of short- and long-range non-linear interactions.
 
\color{black}
\begin{acknowledgments}
\noindent For partial support, S.S. and L.T. thank Funda\c c\~ao de Amparo 
\`a Pesquisa do Estado de S\~ao Paulo [Contracts 2020/02185-1, 2017/05660-0, 2024/04174-8, 2024/01533-7].  
L.T. also acknowledges partial support from Conselho Nacional de Desenvolvimento 
Cient\'\i fico e Tecnol\'ogico (Proc. 304469-2019-0). R.R. wishes 
to acknowledge the financial assistance from DST(DST-CURIE-PG/2022/54) and ANRF(formerly DST-SERB) 
(CRG/2023/008153). 
\end{acknowledgments}

\color{black}

\end{document}